\documentclass[a4paper]{jpconf}
\usepackage{graphicx}
\usepackage{subfig}

\newcommand{\sqrts}{\ensuremath{\sqrt{s}}}
\newcommand{\sqrtsNN}{\ensuremath{\sqrt{s_{\rm{NN}}}}}
\newcommand{\gev}{GeV/$c$}

\newcommand{\Dzero}{\ensuremath{D^{\rm{0}}}}
\newcommand{\jpsi}{\ensuremath{J/\psi}}
\newcommand{\ups}{\ensuremath{\Upsilon}}
\newcommand{\pT}{\ensuremath{p_{\rm{T}}}}
\newcommand{\RAA}{\ensuremath{R_{\rm{AA}}}}
\newcommand{\RpPb}{\ensuremath{R_{\rm{pPb}}}}

\newcommand{\npart}{\ensuremath{N_{\rm{part}}}}

\begin{document}
\title{Experimental summary for heavy flavor production}

\author{Rongrong Ma}

\address{Building 510A, Physics department, Brookhaven National Laboratory, Upton, NY 11973, USA}

\ead{marr@bnl.gov}

\begin{abstract}
Measurements of heavy flavor production in heavy-ion collisions have played an important role in understanding the properties of the quark-gluon plasma created in such collision. Due to their large masses, heavy flavor quarks present unique sensitivity to the kinematics as well as the dynamics of the hot and dense medium. In this article, a selection of recent measurements on heavy flavor production in p+p, p+A and A+A collisions at both RHIC and LHC energies will be presented. The measurements in p+p collisions serve as benchmarks to fundamental theories, and as references to similar studies in A+A collisions where the hot medium effects are present. On the other hand, the measurements in p+A collisions can help to quantify the cold nuclear matter effects which are also in effect  in A+A collisions and thus need to be taken into account when interpreting the measurements in heavy-ion collisions. The experimental results from A+A collisions are discussed and compared to theoretical calculations, which can shed lights on the understanding of the quark-gluon plasma.
\end{abstract}

\section{Introduction}
In relativistic heavy-ion collisions, a new state of matter, usually referred to as the ``Quark-Gluon Plasma (QGP)", has been created in laboratory at RHIC since more than fifteen years ago \cite{Adams:2005dq}. Since then, understanding the properties of such nuclear matter under extreme conditions has been the central focus of the heavy-ion physics. Among various probes, heavy flavor quarks are playing an essential role in studying the kinematics and dynamics of the QGP \cite{Frawley:2008kk,Andronic:2015wma}.  

Open heavy flavor quarks (charm and bottom) are predominately produced during the initial hard scatterings at the early stage of the heavy-ion collisions. They consequently interact with the medium throughout the entire evolution of the medium, and thus encode information from various stages of the nuclear matter. The interaction with the medium leads to substantial energy loss, both radiative and collisional, for heavy quarks \cite{Braaten:1991we,Dokshitzer:2001zm}. By studying the dependence of energy loss on collision setup, e.g. energy, geometry, system, etc, as well as comparing the measurements to that of light hadrons and among different heavy flavor hadrons, one can better understand the properties of the QGP and extract key parameters for the plasma \cite{Averbeck:2013oga}.

Quarkonium is another probe sensitive to the partonic nature of the QGP. Dissociation of the \jpsi\ meson in the medium due to the color-screening of the potential between charm and anti-charm quarks by the surrounding partons was proposed as an evidence for the formation of the deconfined nuclear matter \cite{Matsui:1986dk}. However, regeneration of charm and anti-charm quarks, abundantly produced in the medium, significantly alters the observed \jpsi\ suppression pattern \cite{Zhou:2014kka,Zhao:2010nk}. Other effects, such as cold nuclear matter effects, co-mover absorption, formation time, feed-down, etc, need to be taken into account as well when interpreting the measurements of \jpsi\ suppression in heavy-ion collisions \cite{Zhao:2010nk}. On the other hand, the \ups\ mesons suffer much less from the regeneration contribution due to the much smaller cross-section for bottom quark production in the medium compared to charm quarks. The three different \ups\ states are supposed to dissociate at different temperatures due to different binding energies, and measurements of this sequential suppression can help deduce the temperature of the QGP \cite{Mocsy:2007yj}. 

\section{Heavy flavor production in p+p and p+A collisions}
\begin{figure}[h]
\begin{minipage}{18pc}
\begin{center}
\includegraphics[width=16pc]{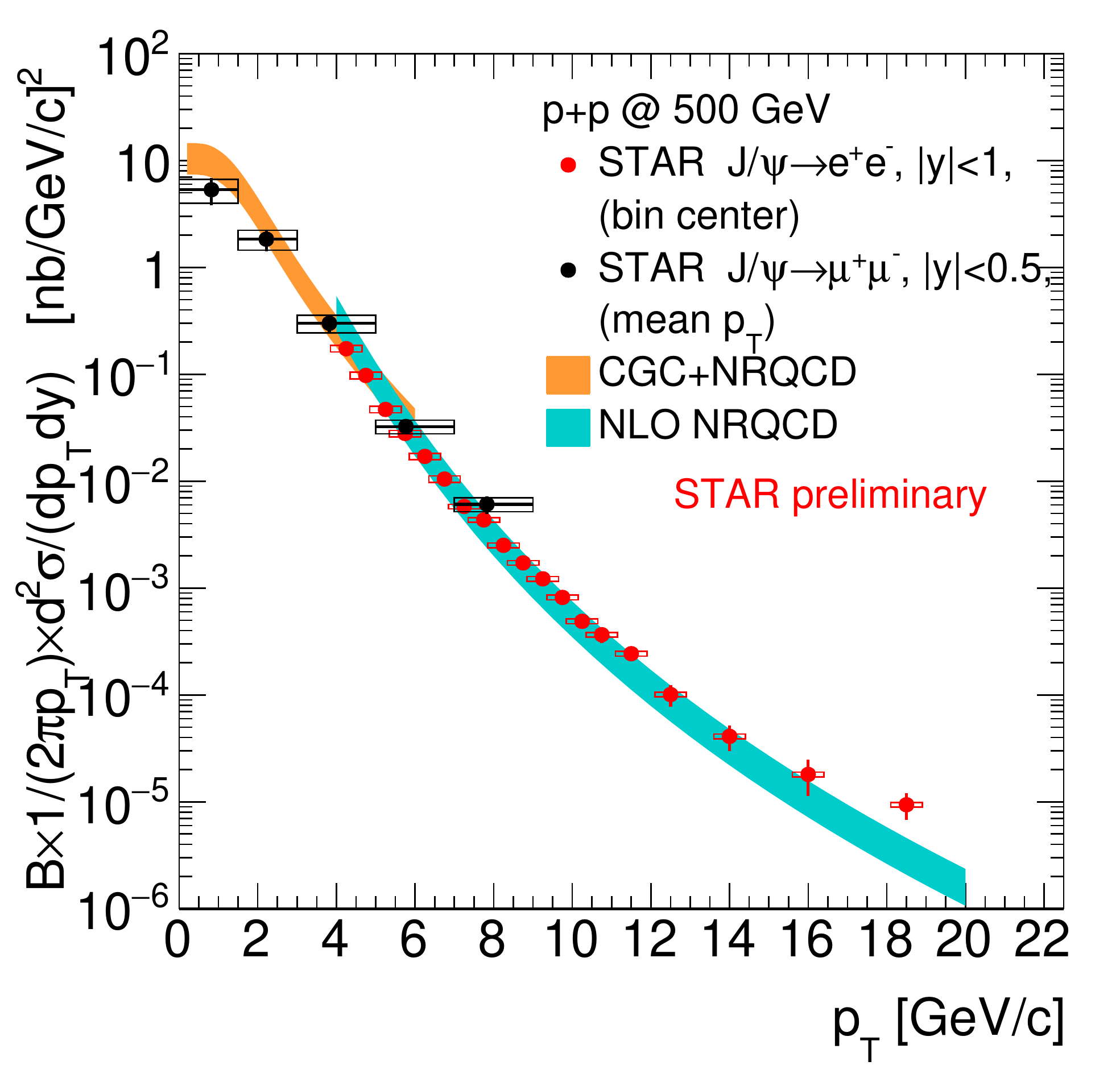}
\caption{\label{Fig_STAR_pp500_Jpsi}\jpsi\ cross-section in p+p collisions at \sqrts\ = 500 GeV measured by the STAR experiment for $0 < \pT < 20$ \gev.}
\end{center}
\end{minipage}
\hspace{1pc}%
\begin{minipage}{18pc}
\begin{center}
\includegraphics[width=16pc]{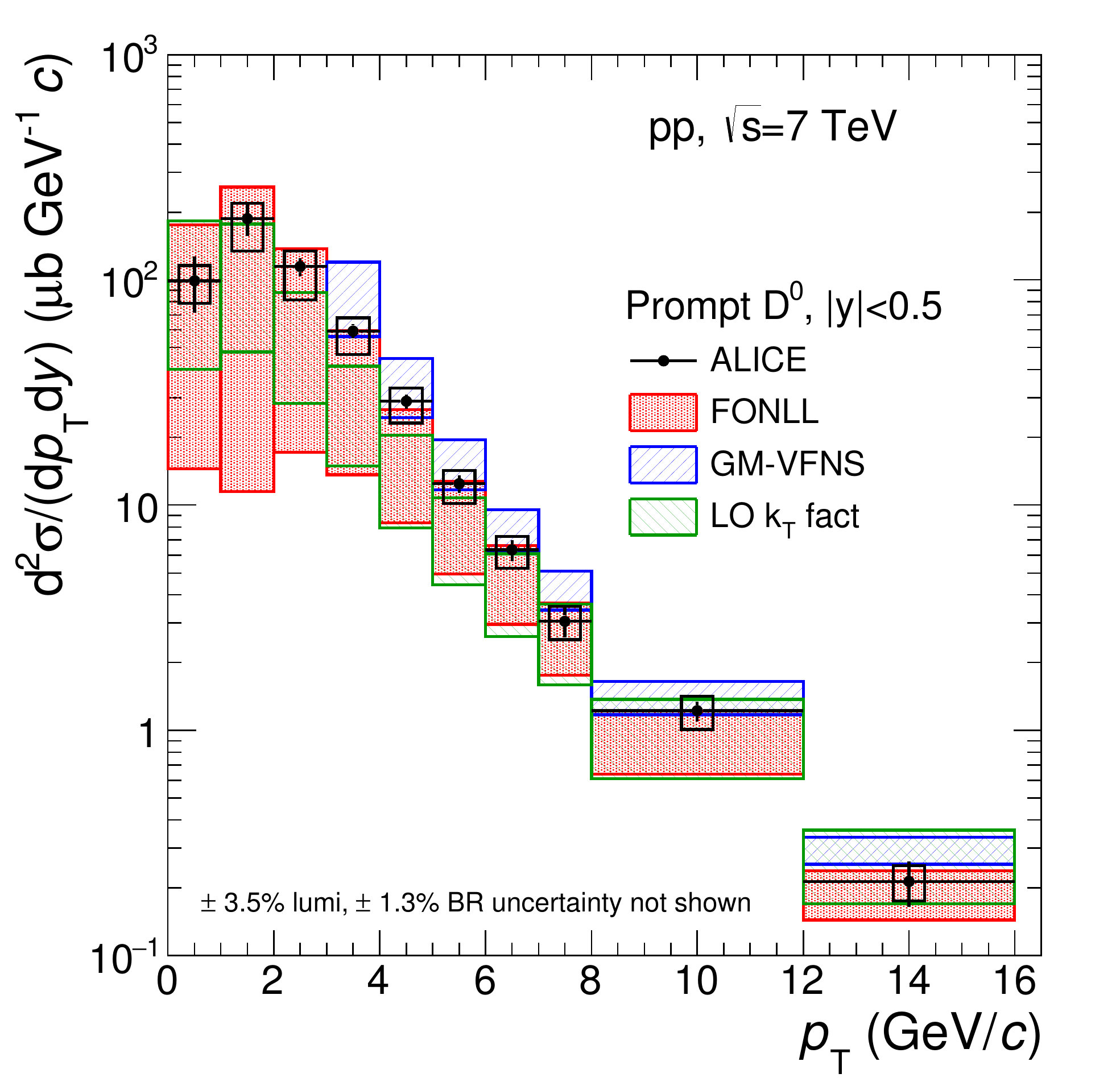}
\caption{\label{Fig_ALICE_pp7_D}\Dzero\ cross-section in p+p collisions at \sqrts\ = 7 TeV measured by the ALICE experiment for $0 < \pT < 16$ \gev\ \cite{Adam:2016ich}.}
\end{center}
\end{minipage} 
\end{figure}

Due to the large masses, the heavy flavor production can be calculated analytically using perturbative-QCD (pQCD), which makes it a well-controlled probe to the QGP. The differential cross-section of \jpsi\ in p+p collisions at \sqrts\ = 500 GeV measured by the STAR experiment for $0 < \pT < 20$ \gev\ is shown in Fig. \ref{Fig_STAR_pp500_Jpsi}.
A similar measurement for \Dzero\ meson in p+p collisions at \sqrts\ = 7 TeV by the ALICE experiment is shown in Fig. \ref{Fig_ALICE_pp7_D} covering the range of $0 < \pT < 16$ \gev\ \cite{Adam:2016ich}. For both cases, QCD-inspired models can describe the data fairly well within the  uncertainties from both model calculations and experimental measurements \cite{Adam:2016ich,Ma:2010yw,Ma:2014mri}.

\begin{figure}[h]
\begin{center}
\subfloat[]{\includegraphics[width=0.37\textwidth]{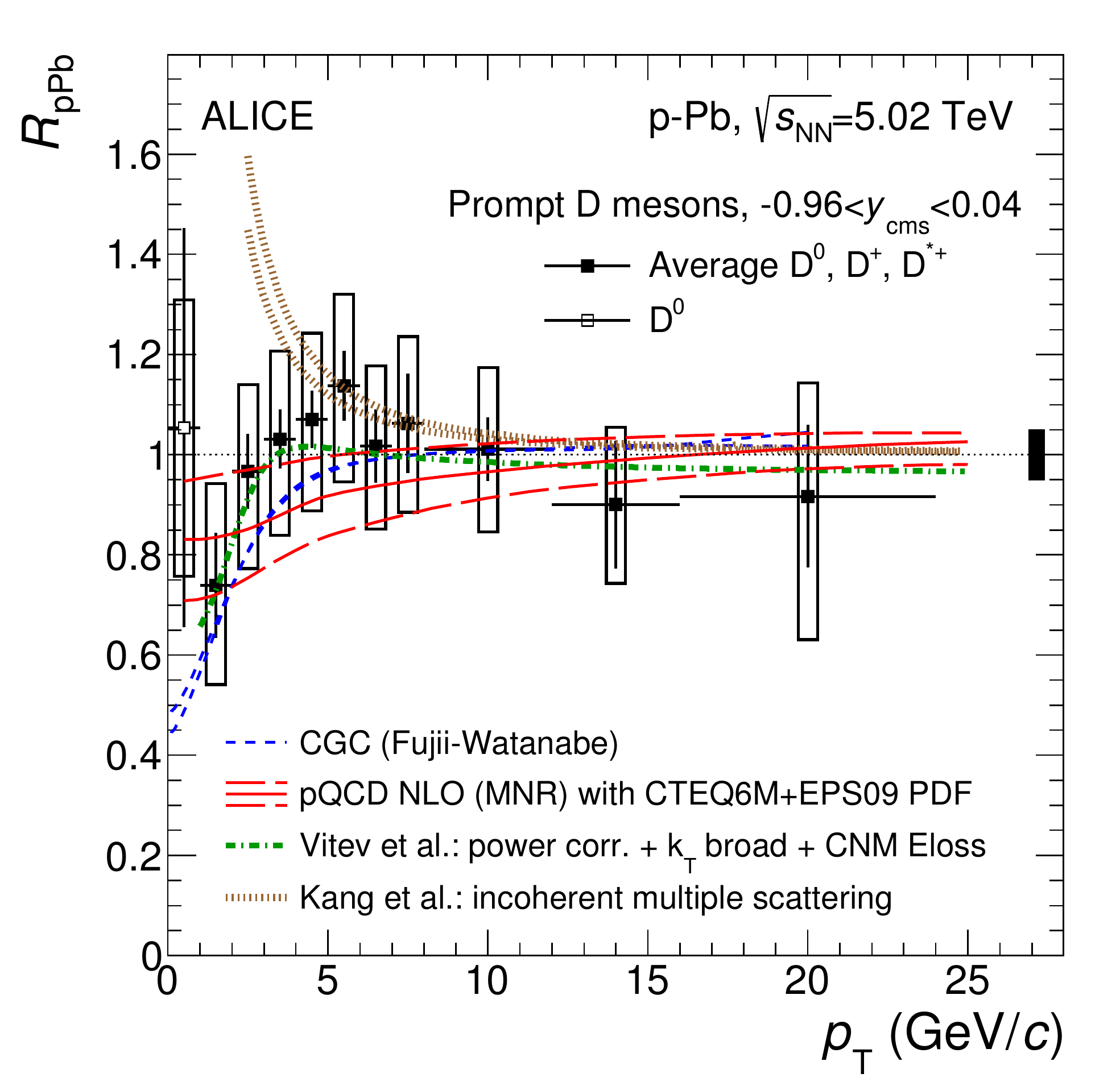}\label{Fig_ALICE_pPb_DRpPb}}
\hspace{2pc}%
\subfloat[]{\includegraphics[width=0.55\textwidth]{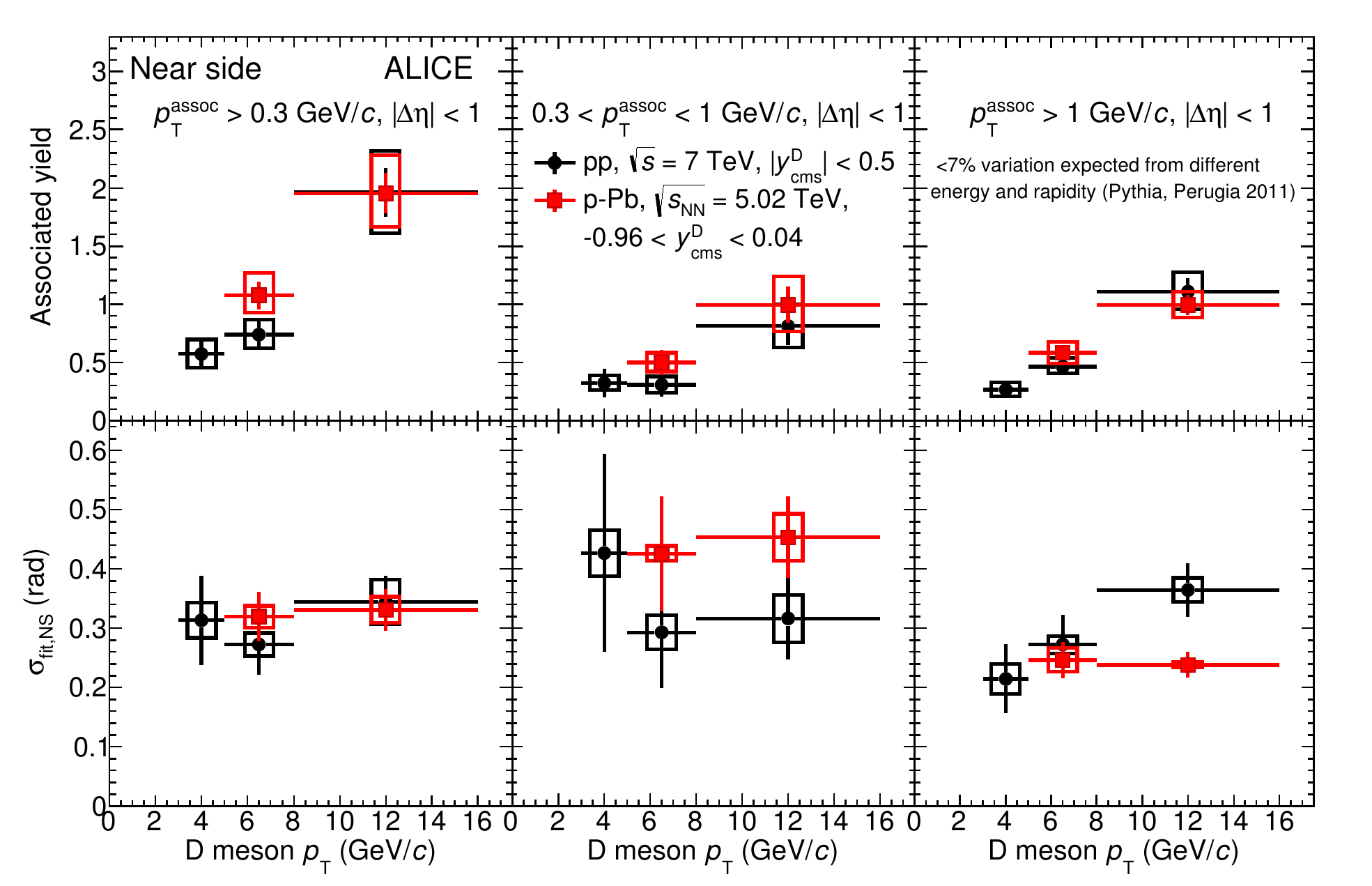}\label{Fig_ALICE_pPb_Dh}}
\caption{Measurements of \Dzero\ meson production in minimum-bias p+Pb collisions at \sqrtsNN\ = 5.02 TeV by the ALICE experiment. (a) \Dzero\ \RpPb\ as a function of \pT\ along with model calculations \cite{Adam:2016ich}. (b) Yields of associated hadrons (upper) and the widths of the correlation peaks (lower) extracted from \Dzero-hadron correlations on the near-side (squares). Similar measurements in p+p collisions at \sqrts\ = 7 TeV (circles) are shown for comparison \cite{ALICE:2016clc}.}
\end{center}
\end{figure}

Figure \ref{Fig_ALICE_pPb_DRpPb} shows the measurements of nuclear modification factors (\RAA) for \Dzero\ mesons as a function of \pT\ in minimum-bias p+Pb collisions at \sqrtsNN\ = 5.02 TeV by the ALICE collaboration \cite{Adam:2016ich}. \RpPb\ is consistent with unity at high \pT\ and shows a hint of nuclear showing at low \pT. Model calculations incorporating only Cold Nuclear Matter (CNM) effects agree with data reasonably well within the large uncertainties \cite{Adam:2016ich}. The ALICE collaboration also measured \Dzero-hadron correlations. The yields of associated hadrons (upper) and the widths of the correlation peaks (lower) as a function of \Dzero\ \pT\ on the near-side are shown for both p+p collisions at \sqrts\ = 7 TeV (circles) and p+Pb collisions at \sqrtsNN\ = 5.02 TeV (squares) in Fig. \ref{Fig_ALICE_pPb_Dh} \cite{ALICE:2016clc}. They are seen to be consistent indicating that the fragmentation processes to \Dzero\ mesons are largely unaltered in p+Pb collisions. 

\begin{figure}[h]
\begin{center}
\subfloat[\jpsi\ \RpPb\ as a function of rapidity by the LHCb experiment \cite{Aaij:2016eyl}.]{\includegraphics[width=0.45\textwidth]{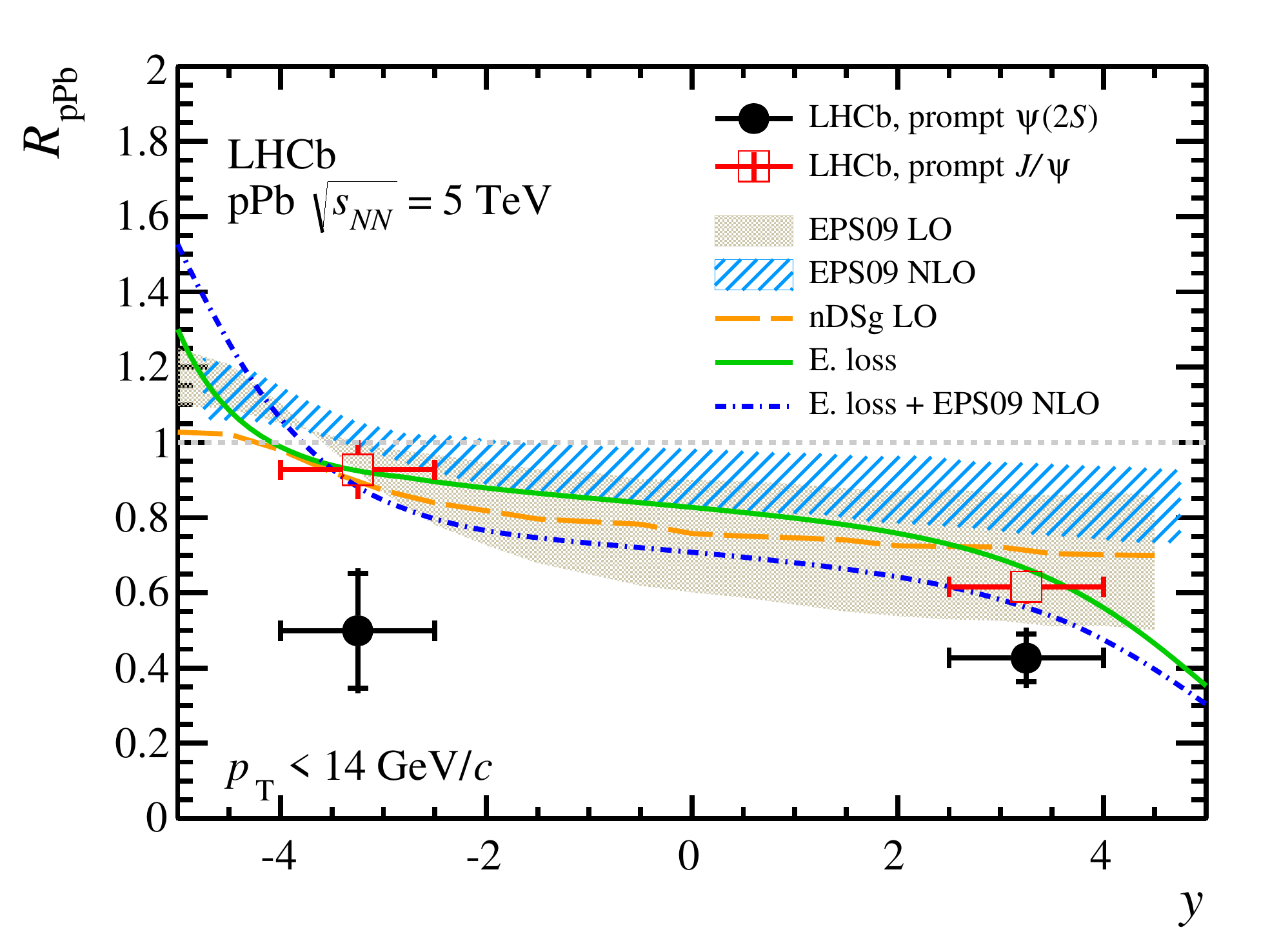}\label{Fig_LHCb_pPb_Jpsi}}
\hspace{2pc}%
\subfloat[$\Upsilon(1\rm{S})$ \RpPb\ as a function of rapidity by the ALICE experiment \cite{Abelev:2014oea}.]{\includegraphics[width=0.45\textwidth]{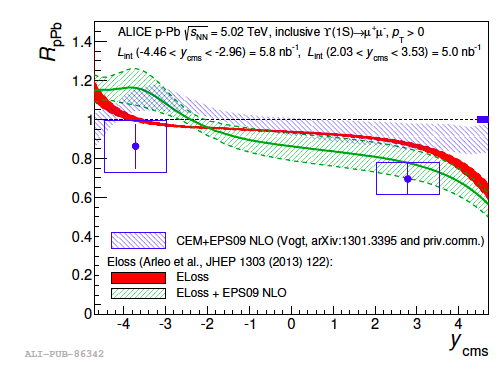}\label{Fig_ALICE_pPb_Ups}}
\caption{Measurements of quarkonium production in minimum-bias p+Pb collisions at \sqrtsNN\ = 5.02 TeV.}
\end{center}
\end{figure}

The modifications to \jpsi\ and \ups\ productions in minimum-bias p+Pb collisions at \sqrtsNN\ = 5.02 TeV are measured as a function of rapidity, as shown in Fig. \ref{Fig_LHCb_pPb_Jpsi} and Fig. \ref{Fig_ALICE_pPb_Ups}, respectively. They are significantly suppressed at forward rapidity while remain nearly unchanged at backward rapidity. This is consistent with nuclear shadowing, but there seems also room left for additional CNM effects, such as energy loss.

\section{Heavy flavor production in A+A collisions}
\subsection{Open heavy flavor}
\begin{figure}[h]
\begin{center}
\subfloat[Au+Au collisions at \sqrtsNN\ = 200 GeV by STAR \cite{Adamczyk:2014uip}.]{\includegraphics[width=0.31\textwidth]{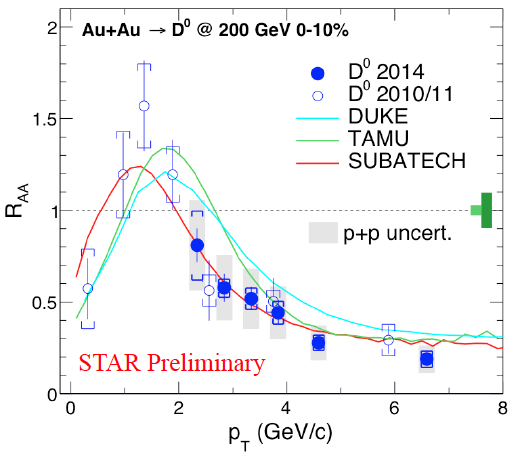}\label{Fig_STAR_AuAu_D0Raa}}
\hspace{1pc}%
\subfloat[Pb+Pb collisions at \sqrtsNN\ = 2.76 TeV by ALICE \cite{Adam:2015sza}.]{\includegraphics[width=0.29\textwidth]{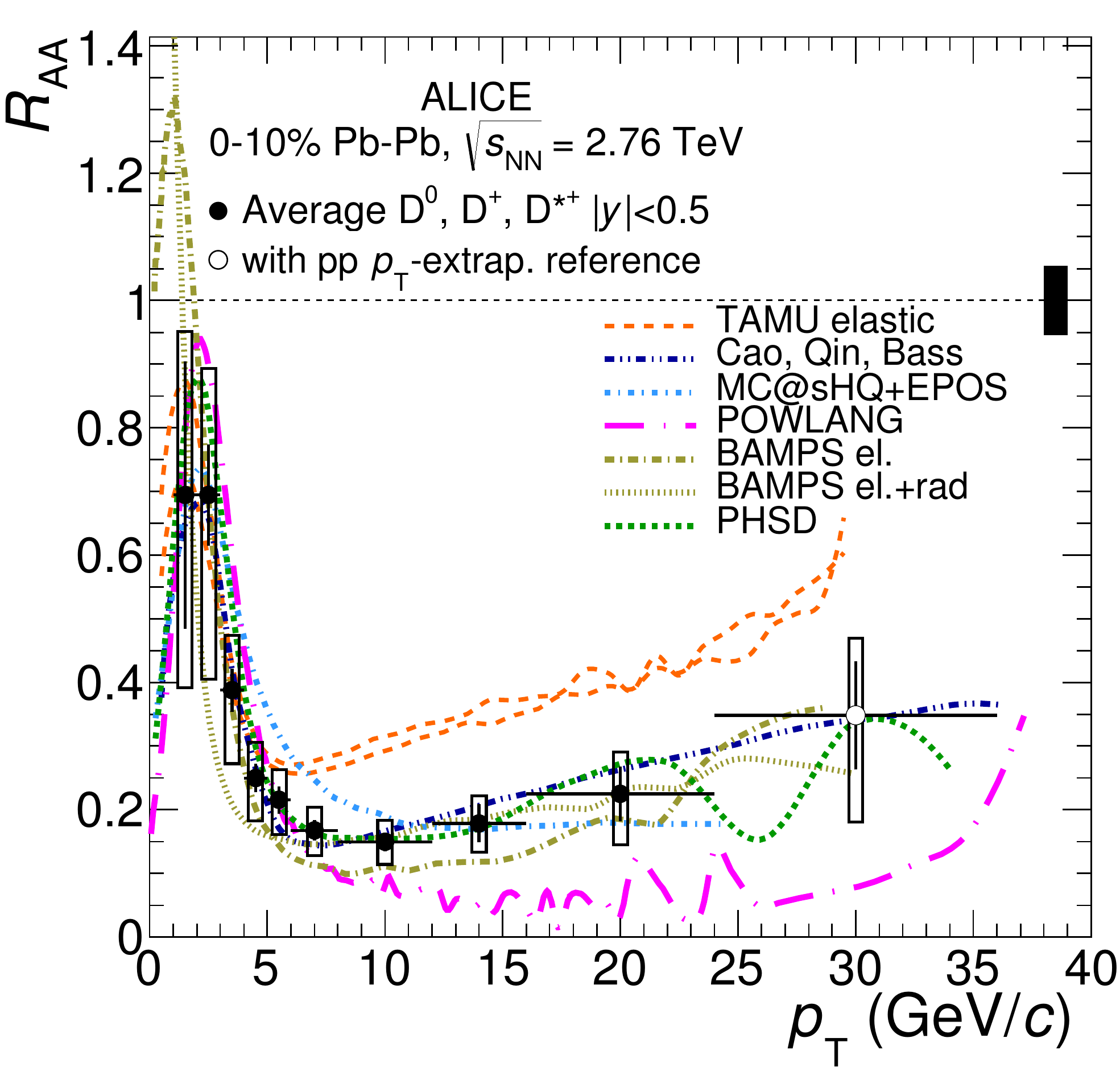}\label{Fig_ALICE_PbPb_D0Raa}}
\hspace{1pc}%
\subfloat[Pb+Pb collisions at \sqrtsNN\ = 5.02 TeV by CMS.]{\includegraphics[width=0.3\textwidth]{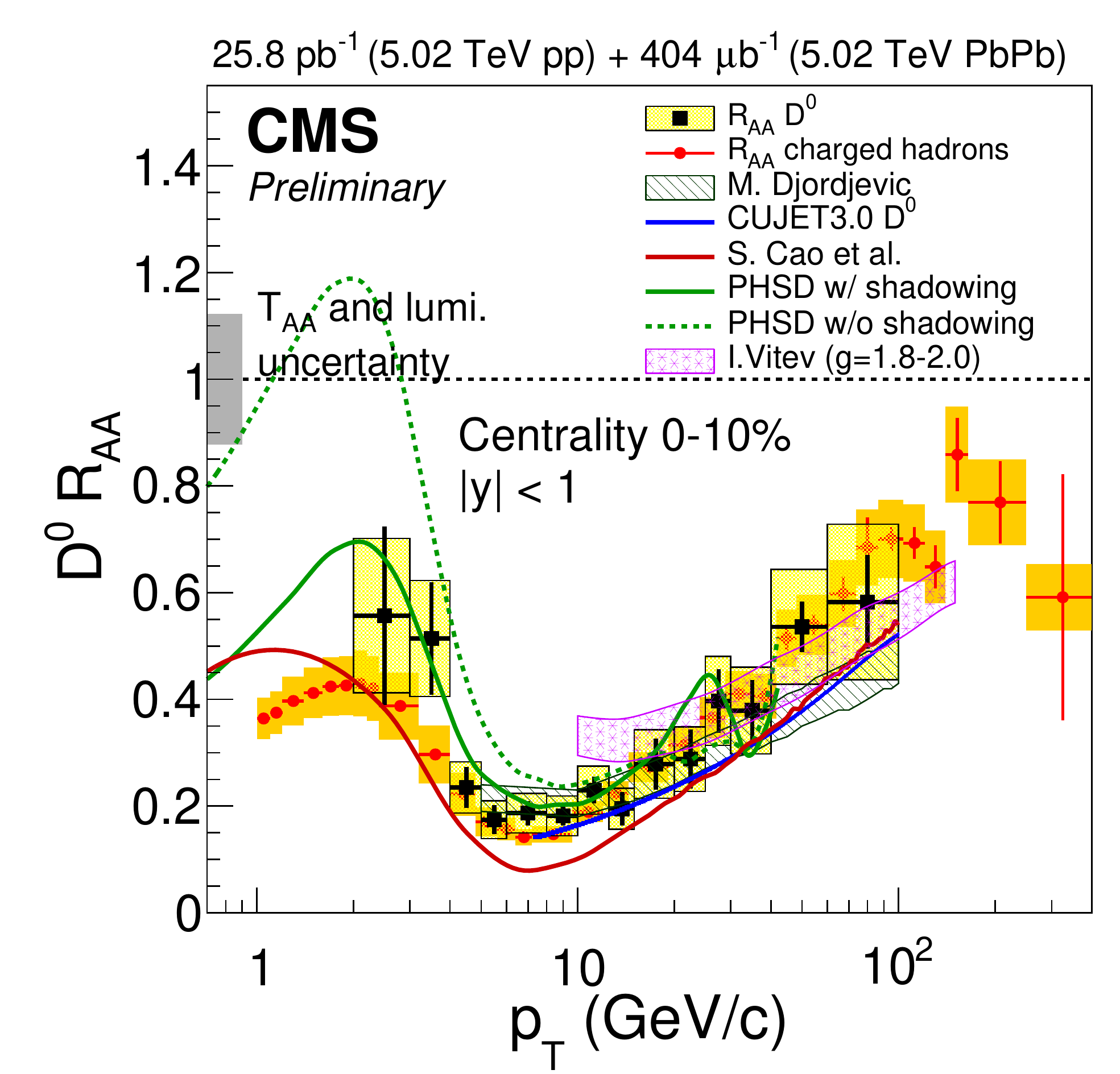}\label{Fig_CMS_PbPb_D0Raa}}
\caption{\label{D0Raa}Measurements of \Dzero\ \RAA\ in 0-10\% central heavy-ion collisions at RHIC and LHC.}
\end{center}
\end{figure}

Figure \ref{D0Raa} shows measurements of \Dzero\ \RAA\ as a function of \pT\ in 0-10\% central heavy-ion collisions at RHIC and LHC energies \cite{Adamczyk:2014uip,Adam:2015sza}. Compared to scaled p+p reference, \Dzero\ production is significantly suppressed, up to a factor of 5-6, at high \pT, indicating strong interactions between charm quarks and the hot medium. A bump structure at lower \pT\ is observed at all three collision energies, which could be explained by the coalescence of charm quarks with the light quarks in the medium. Unlike the case at RHIC, the bump structure at the LHC energies does not exceed unity, probably because of the stronger shadowing effects. Several models incorporating energy loss mechanisms differently can qualitatively describe the data.

\begin{figure}[h]
\begin{center}
\subfloat[]{\includegraphics[width=0.4\textwidth]{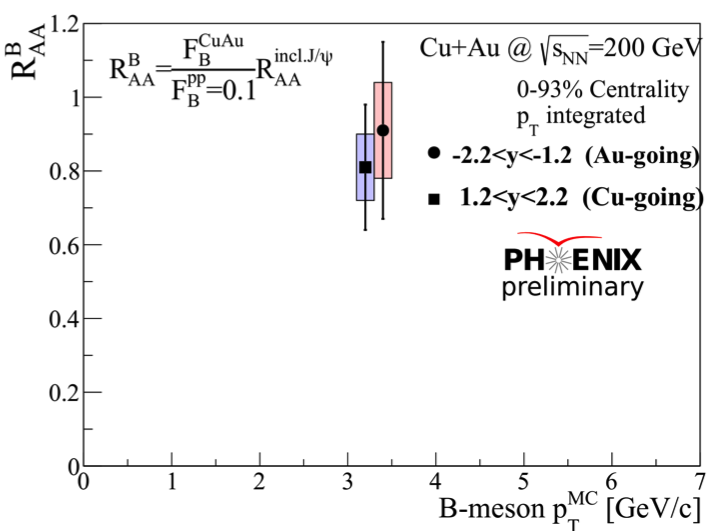}\label{Fig_PHENIX_CuAu_NonJpsi}}
\hspace{4pc}%
\subfloat[]{\includegraphics[width=0.35\textwidth]{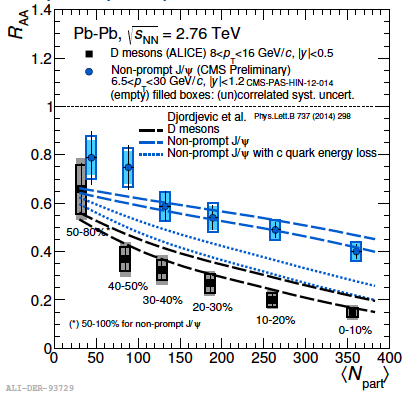}\label{Fig_ALICE_PbPb_NonJpsi}}
\caption{\label{NonJpsiRaa}(a) B-meson \RAA\ extracted from measurements of non-prompt \jpsi\ in Cu+Au collisions at \sqrtsNN\ = 200 GeV from the PHENIX experiment. (b) \RAA\ of \Dzero\ and non-prompt \jpsi\ as a function of \npart\ \cite{Miller:2007ri} in Pb+Pb collisions at \sqrtsNN\ = 2.76 TeV \cite{Adam:2015nna}.}
\end{center}
\end{figure}

The production of non-prompt \jpsi\ from B-meson decays at forward rapidities is measured in minimum-bias Cu+Au collisions at \sqrtsNN\ = 200 GeV by the PHENIX experiment. With the help of Monte Carlo models for decay kinematics, the B-meson \RAA\ is extracted as a function of B-meson \pT\ as shown in Fig. \ref{Fig_PHENIX_CuAu_NonJpsi}, which is seen to be systematically higher than \Dzero\ \RAA\ at the same \pT. 
Such a comparison is also done in Fig. \ref{Fig_ALICE_PbPb_NonJpsi} for Pb+Pb collisions at \sqrtsNN\ = 2.76 TeV at the LHC \cite{Adam:2015nna}, and similarly the \RAA\ of non-prompt \jpsi\ is significantly higher than that of \Dzero\ mesons. Model calculations \cite{Djordjevic:2014tka} taking into account both the expected mass hierarchy of the radiative energy loss, i.e. $\Delta E_{c} > \Delta E_{b} $, as well as spectrum shape can describe the data fairly well, as shown in Fig. \ref{Fig_ALICE_PbPb_NonJpsi}. Other calculations, such as the MC$@$sHQ+EPOS2 model \cite{Nahrgang:2013xaa} and the TAMU elastic model \cite{He:2012xz}, are also compatible with such a hierarchy.

\subsection{Quarkonium}
\begin{figure}[h]
\begin{center}
\subfloat[\jpsi\ at mid-rapidity in Au+Au collisions at \sqrtsNN\ = 200 GeV and Pb+Pb collisions at \sqrtsNN\ = 2.76 TeV.]{\includegraphics[width=0.45\textwidth]{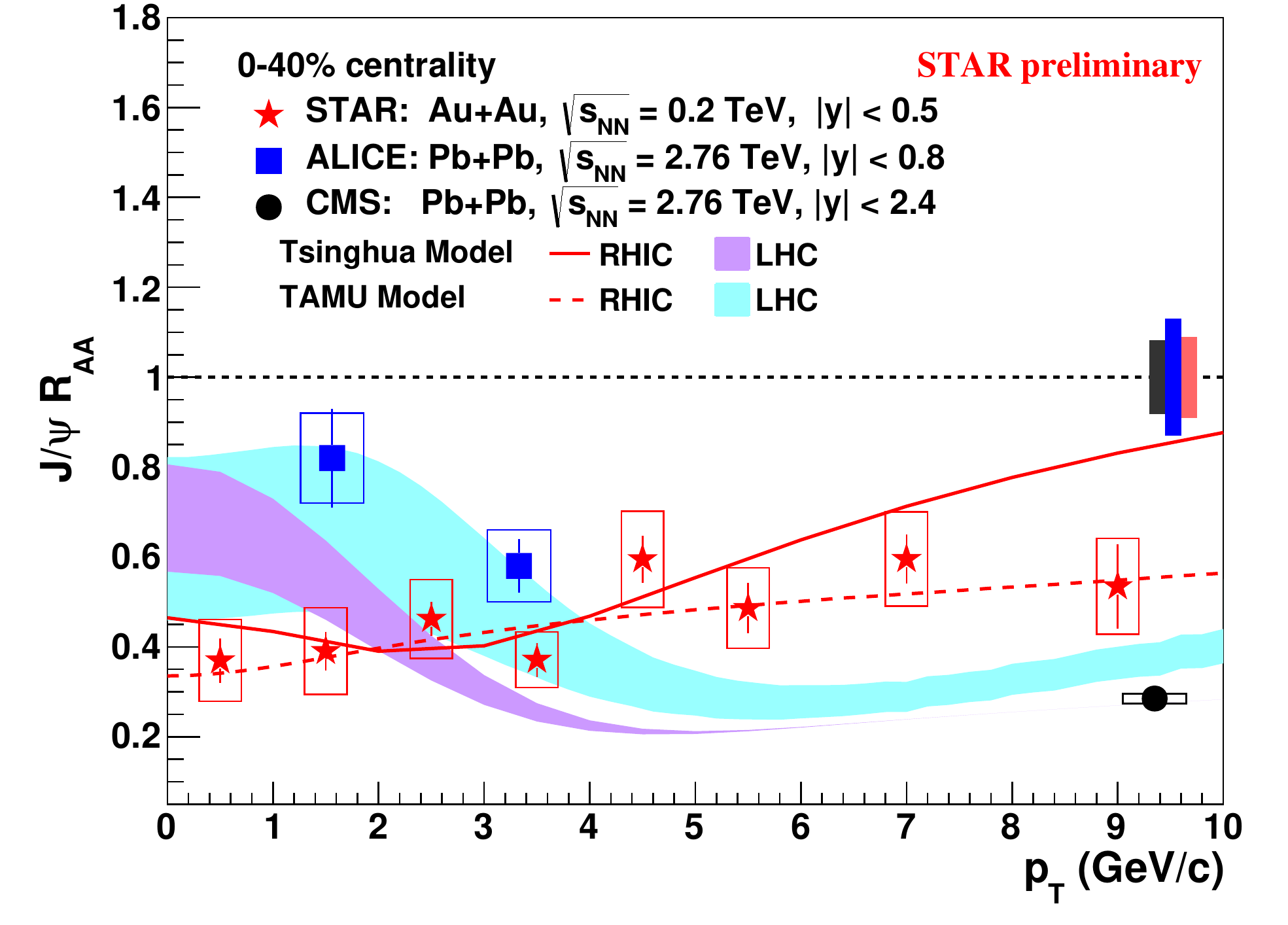}\label{Fig_STAR_AuAu_JPsi}}
\hspace{2pc}%
\subfloat[\jpsi\ at forward-rapidity in Pb+Pb collisions at \sqrtsNN\ = 2.76 and 5.02 TeV \cite{Adam:2016rdg}.]{\includegraphics[width=0.45\textwidth]{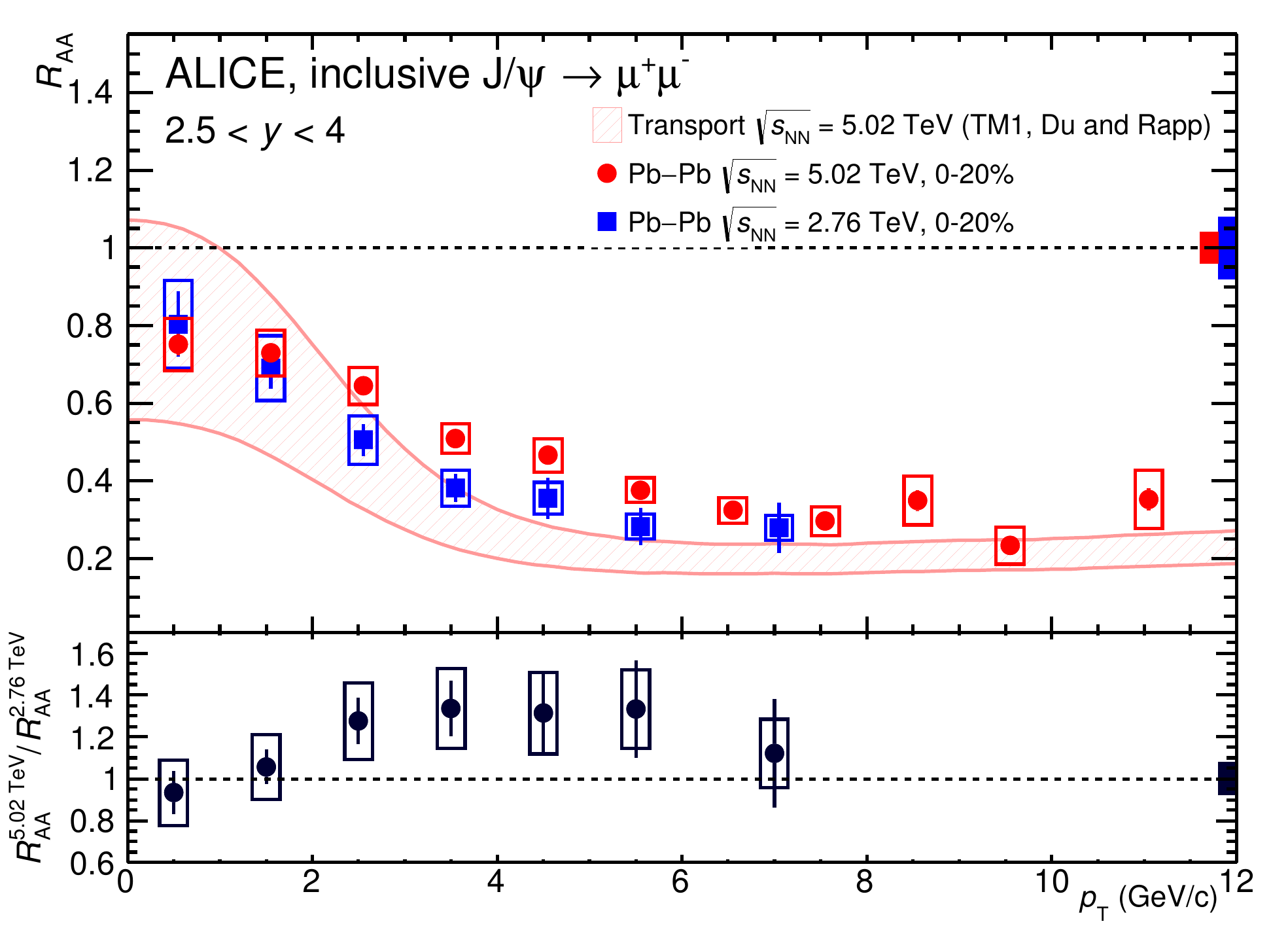}\label{Fig_ALICE_PbPb_502_Jpsi}}
\caption{\label{JpsiRaa}Measurements of \jpsi\ \RAA\ as a function of \pT\ at RHIC and LHC energies.}
\end{center}
\end{figure}

A first measurement of \jpsi\ suppression within $0 < \pT < 10$ \gev\ at mid-rapidity via the di-muon channel is done by the STAR experiment for 0-40\% central Au+Au collisions at \sqrtsNN\ = 200 GeV, as shown in Fig. \ref{Fig_STAR_AuAu_JPsi}, along with similar measurements in Pb+Pb collisions at \sqrtsNN\ = 2.76 TeV \cite{Adam:2015rba,Chatrchyan:2012np}. At low \pT, the \jpsi\ \RAA\ is higher at the LHC which can be explained by a larger regeneration contribution due to the larger cross-section for charm quark production, while at high \pT\ the \jpsi\ \RAA\ is lower at the LHC due probably to the larger dissociation rate because of the higher medium temperature.
Similar measurements at forward rapidity are shown in Fig. \ref{Fig_ALICE_PbPb_502_Jpsi} for 0-20\% central Pb+Pb collisions at \sqrtsNN\ = 5.02 TeV \cite{Adam:2016rdg}, which are seen to be systematically higher than that at \sqrtsNN\ = 2.76 TeV for $2 < \pT < 6$ \gev. One possible explanation is that the regeneration contribution dominates the dissociation in this kinematic range. Transport models from Tsinghua \cite{Zhou:2014kka} and TAMU \cite{Zhao:2010nk} groups include both regeneration and dissociation contributions, and can qualitatively describe the data.

Suffering less from regeneration contribution compared to \jpsi, \ups\ mesons are considered a cleaner probe.
\begin{figure}[h]
\begin{center}
\subfloat[]{\includegraphics[width=0.45\textwidth]{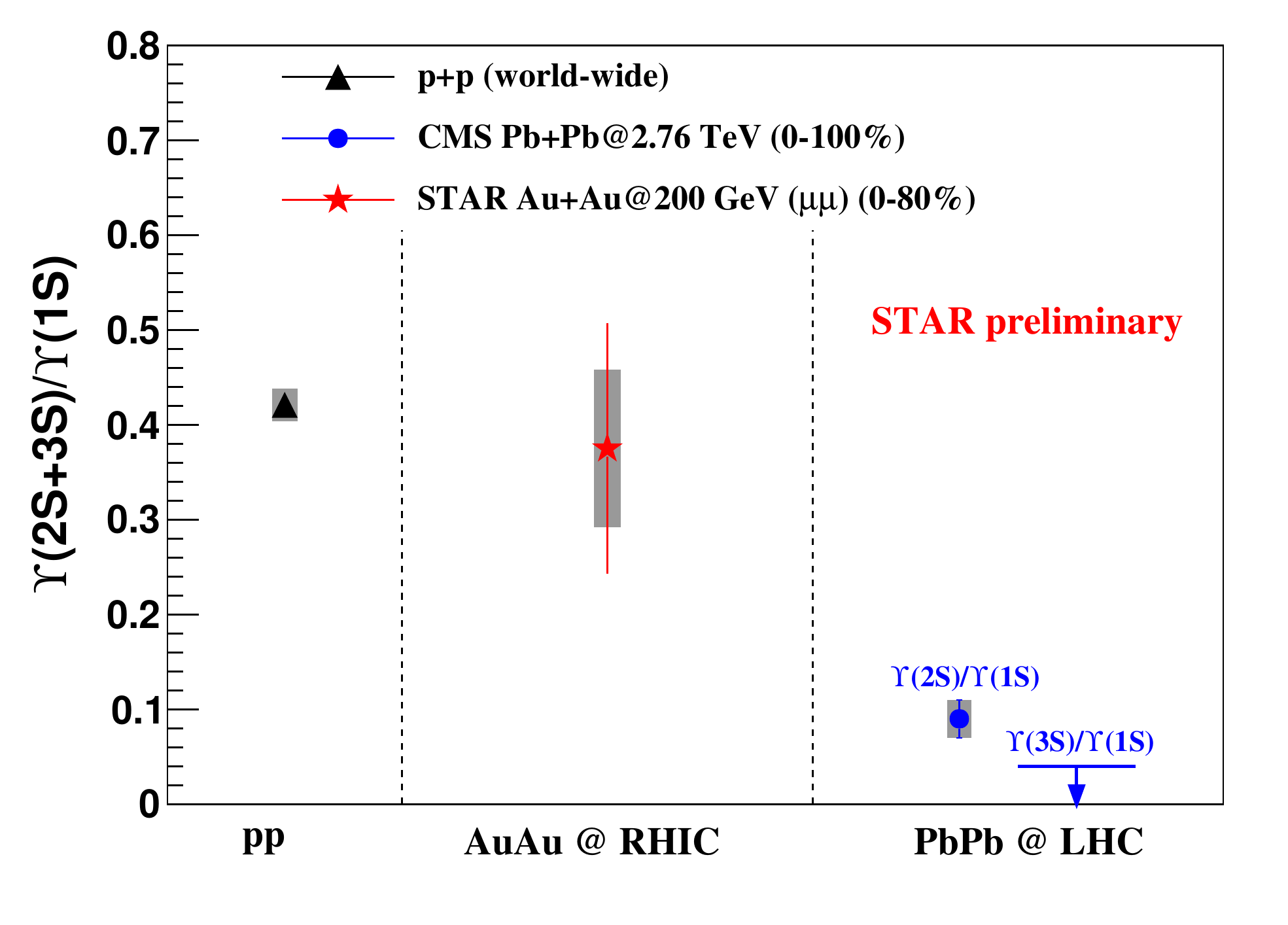}\label{Fig_STAR_AuAu_Ups}}
\hspace{2pc}%
\subfloat[]{\includegraphics[width=0.45\textwidth]{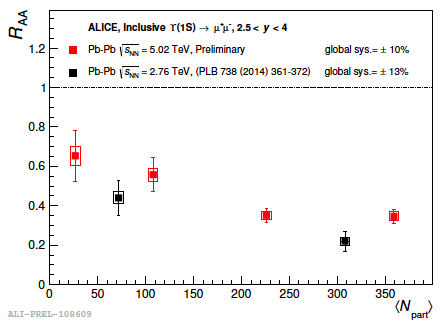}\label{Fig_ALICE_PbPb_502_Ups}}
\caption{\label{UpsilonRaa}(a) Measurements of ratio of excited \ups\ states to ground state in Au+Au collisions at \sqrtsNN\ = 200 GeV, compared with that from world-wide p+p data \cite{Zha:2013uoa} and in Pb+Pb collisions at \sqrtsNN\ = 2.76 TeV. (b) Measurements of \ups\ \RAA\ as a function of \npart\ in Pb+Pb collisions at \sqrtsNN\ = 2.76 and 5.02 TeV.}
\end{center}
\end{figure}
Figure \ref{Fig_STAR_AuAu_Ups} shows the ratio of $\Upsilon(2\rm{S}+3\rm{S})/\Upsilon(1\rm{S})$ measured in Au+Au collisions at \sqrtsNN\ = 200 GeV. Compared to similar measurements at higher collision energies at the LHC, there is a hint of less relative suppression of excited \ups\ states to the ground state at RHIC. \ups\ \RAA\ as a function of \npart\ are shown in Fig. \ref{Fig_ALICE_PbPb_502_Ups} for Pb+Pb collisions at \sqrtsNN\ = 2.76 and 5.02 TeV, where the larger \RAA\ values for 5.02 TeV collisions could be due to a larger regeneration contribution as well since the bottom quark production becomes sizable at such a high collision energy. Future measurements of feed-down contributions to the \ups\ states in the heavy-ion collisions are critical for a thorough understanding of different mechanisms in play.

\section{Summary}
Measurements of heavy flavor production in heavy-ion collisions have revealed many important aspects of the QGP properties in the last two decades. The strong interaction of heavy quarks with the medium leads to considerable energy loss for both charm and bottom quarks. The suppression seems to be larger for \Dzero\ meson compared to non-prompt \jpsi\ coming from  B-meson decays, which is consistent with the expected mass hierarchy \cite{Dokshitzer:2001zm}. These measurements can help constrain the transport coefficient ($\hat{q}$) of the medium. On the other hand, the color-screening mechanism responsible for quarkonium dissociation in the medium seems to hold at both RHIC and LHC even though an additional contribution from regeneration also plays a critical role in interpreting the suppression patterns at different collision energies. This provides potentially a strong evidence for the QGP formation. With increased luminosity at RHIC and new beam energies at LHC, heavy flavor measurements will continue to contribute significantly towards a comprehensive understanding of the QGP properties.

\section*{References}
\bibliography{RM_sQM2016}

\end{document}